\documentclass[a4paper,fleqn,usenatbib]{mnras}

\usepackage{newtxtext,newtxmath}

\usepackage[T1]{fontenc}
\usepackage{ae,aecompl}

\usepackage{graphicx}	
\usepackage{amsmath}	
\usepackage{amssymb}	

\title[Orbital stability in the Solar System]{Orbital stability in the Solar System for arbitrary inclinations and eccentricities: planetary perturbations versus resonances}

\author[T. Gallardo]{
Tabar\'e Gallardo$^{1}$\thanks{E-mail: gallardo@fisica.edu.uy}
\\
$^{1}$Instituto de F\'{i}sica, Facultad
de Ciencias, UdelaR, Igu\'{a} 4225, 11400 Montevideo, Uruguay
}

\date{Accepted XXX. Received YYY; in original form ZZZ}

\pubyear{2019}

\begin{document}
\label{firstpage}
\pagerange{\pageref{firstpage}--\pageref{lastpage}}
\maketitle

\begin{abstract}
Applying the technique of dynamical maps we study the  orbital stability of test particles in the Solar System in the space $(a,e,i)$
defined by $0.1<a<38$ au, $0<e<0.9$ and $0\degr<i<180\degr$ identifying the unstable and stable regions. We find stable niches where small bodies can survive even for very high eccentricities. Mean motion resonances play a fundamental role providing stability against the planetary perturbations specially for high inclination orbits. A stability stripe around $i\sim 150\degr$ is present all along the Solar System. 
We found that the population of objects with semimajor axes between 10 and 30 au is evolving inside a highly unstable region according to our maps. For the inner Solar System we found that
the region between the Hildas and Jupiter is more stable for high eccentricity orbits than for low eccentricity ones.
\end{abstract}




\section{Introduction}
\label{intro}

In  recent articles it was clearly established that the capture in high inclination mean motion resonances (MMRs) is a common phenomena even for retrograde orbits \citep{2013MNRAS.436L..30M,2015MNRAS.446.1998N,2016MNRAS.461.3075F,2017Natur.543..687W,2017MNRAS.472L...1M,2017MNRAS.467.2673N,FERNANDEZ20186}. One decisive factor for the occurrence of these captures is the relatively large strength that some resonances have for the full interval of orbital inclinations \citep{2019Icar..317..121G}. 
However, the capture and stay in resonance  depends not only on the resonance's strength  but also on the planetary perturbations that the particle needs to overcome.
The whole planetary system perturbs the orbits  with different effects according to the small body's orbital elements.  
The aim of this work is to present a global view of the effects of the planetary perturbations in the region between  $0.1 < a <38$ au for the full range of orbital eccentricities and 
inclinations. This purely numerical study is representative for a relatively short time scale of about 1000 orbital revolutions of the test particle. Dynamical evolutions that appear in much longer time scales will not be reflected in this study.

\begin{figure}
	\centering
	\includegraphics[width=1.0\linewidth]{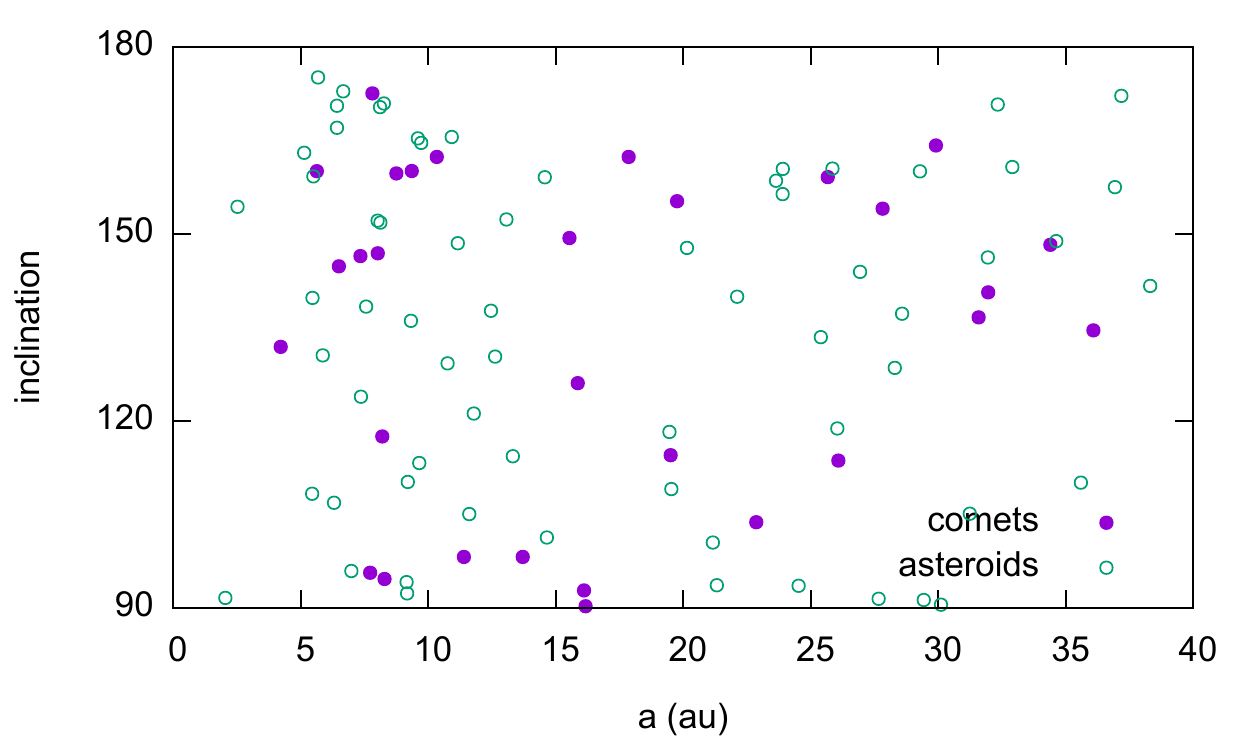}
	\caption{Observed population of retrograde asteroids and comets with $a<40$ au. 	Source:  MPC, February 2019.}
	\label{obsretro}
\end{figure}

The particularity of this study is that it covers all inclinations from 
$0\degr$ to $180\degr$. The effects of the planetary perturbations on orbits with arbitrary inclinations  were intensively studied for the particular case of quasi-parabolic comets \citep[see for example][]{2005ASSL..328.....F}.
On the other hand, 
several studies have been done analyzing the stability of orbits in the plane $(a,e)$  assuming small or zero orbital inclinations.
For example, \citet{1989AJ.....98.1477T} and
\cite{1990Natur.345...49T} studied the stability of the trans-Neptunian region by means of Lyapunov exponents for diverse eccentricities but low inclinations. 
Also, in \citet{1990AJ....100.1680G}  the orbital evolution of initial circular orbits with $i=0\degr$ and $i=10\degr$ with respect to the invariable plane was studied in the outer solar system.
The cornerstone paper by \cite{1993AJ....105.1987H} showed the stability of initial circular and zero inclination orbits for a wide region from 5 to 50 au. 
Some stability explorations in inclinations were done for the trans-Neptunian region:
we can mention \citet{1995AJ....110.3073D}, and in particular
the systematic survey performed by
\citet{1999Icar..140..341G} and \citet{1999Icar..140..353G}
of the stability of particles with initial $4.6<a<30$ au with random eccentricities  and $i<90\degr$, but strongly concentrated near the ecliptic.
The inner solar system for circular and very low inclination orbits was studied by
\citet{1999Natur.399...41E}.
Some other explorations of the planetary perturbations on test particles have been done since then but always related to low inclination orbits \citep[see for example][]{2001Icar..152....4R}. These works were focused on low inclination orbits because their objective was to understand the dynamical evolution of the early Solar System or the actual distribution of minor bodies which is mostly concentrated near the ecliptic.
However, considering the growing population of retrograde objects (see Figure \ref{obsretro}), it corresponds to take a more detailed look at the stability of orbits in the full range of inclinations.
We hope that our stability maps covering all inclinations will help to 
define the stable and unstable regions in the space $(a,e,i)$ and to
understand whether 
the known high inclination objects are evolving in  stable regions or, on the contrary, they are diffusing inside chaotic regions.
In section \ref{mapsoss} we explain our method, we study the stability in the region of the outer solar system and we analyze the dynamical characteristics of some observed objects, mainly Centaurs. In section \ref{mapsiss} we present the study for the inner planetary system and we focus on the particular population of objects with semimajor axes between the Hildas and Jupiter.  In section   \ref{conc} we summarize the main results.

\section{Dynamical maps for the outer Solar System}
\label{mapsoss}

\begin{figure*}
	\centering
	\includegraphics[width=0.9\linewidth]{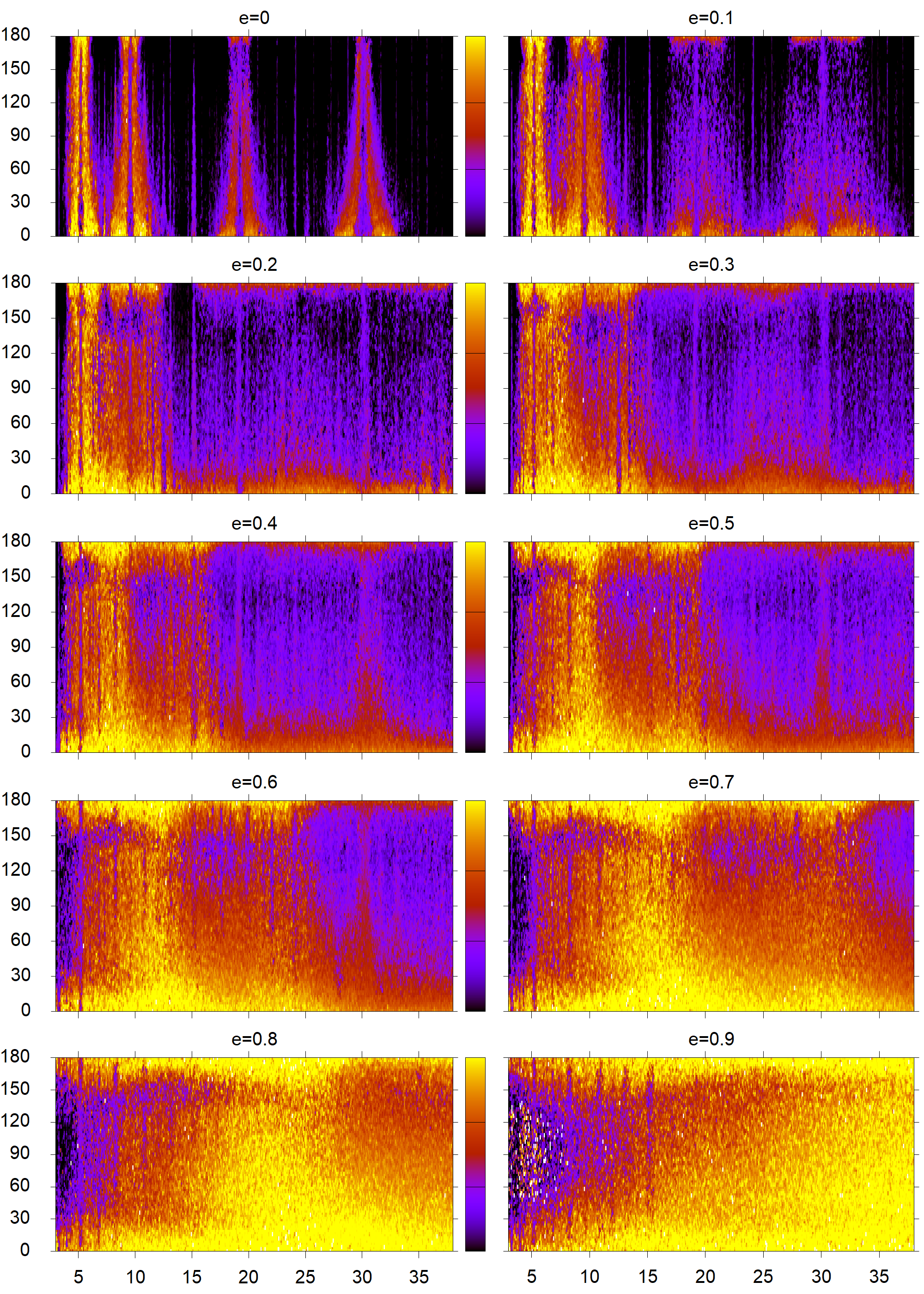}
	\caption{Dynamical maps in terms of initial $(a,i)$ showing the diffusion in $a$ for test particles in the Solar System with initial random $\omega,\Omega,M$ and different initial eccentricities. Horizontal axis is initial $a$ in au and vertical axis is initial inclination in degrees. 
		Going from top (yellow)  to bottom (black) , the ticmarks of the logarithmic colour scale indicate 
			$(\Delta a) /a = 1, 0.1, 0.01$ and $0.001$.		White pixels correspond to ejections. The dark vertical bands surrounded by chaotic regions that can be recognized in several panels are generated by stable MMRs. For the construction of each map, 63000 test particles per 1000 orbital periods were integrated. }
	\label{deltasem338}
\end{figure*}

We constructed the dynamical maps integrating test particles perturbed by the actual system of outer planets from Jupiter to Neptune with initial conditions of the particles uniformly distributed in the interval  $3 < a <38$ au, $0\degr<i<180\degr$ and with particular initial values for $e$ from 0 to 0.9 in steps of 0.1.  
For the initial set $(\omega,\Omega,M)$ we took  random values between
$0\degr$ and  $360\degr$ from an uniform distribution. 
The code used was an adaptation of EVORB \citep{2002Icar..159..358F}. 	
In these integrations the energy of the system remains nearly constant with oscillations of at most 1 part in $10^8$. 
For the construction of each of the ten maps we integrated 63000 test particles.
Each test particle was integrated for 1000 orbital periods  with an output of one particle's orbital period and we calculated the mean orbital elements after 100 orbital revolutions recording the maximum calculated differences  $\Delta a, \Delta e, \Delta i$ of the mean elements. 
Taking the mean instead of the osculating values we eliminate the short period variations and focus on the diffusion 
of the orbital elements in a timescale long enough to detect variations but short enough to avoid the total loss of the memory of the
initial conditions. 
The recorded variations in the orbital elements are slightly different if we calculate the mean over a different number of orbital revolutions, but these differences do not affect the global picture of the maps. 
In this study we have found that the perturbations by the terrestrial planets can be ignored for our purposes in the considered region  accelerating the numerical integrations. 
The particles having close approaches with the planets have  chaotic evolutions so our results will be only indicative of the order of magnitude of the actual orbital changes that the particles experience. On the contrary, the particles without close encounters with the planets will show a very regular evolution that the integrator will reproduce with confidence.
The region $a<5$ au will be studied more properly in the next section. The most relevant results we obtained involve the $\Delta a$ so we
focus on these results.

Figure \ref{deltasem338}  shows the results for $\Delta a$ in logarithmic colour scale. Black regions
correspond to $(\Delta a)/a \leq 0.001$ and yellow regions to $(\Delta a)/a \geq 1$.
So, dark regions represent  zones where 
diffusion in semimajor axis is minimum, which in principle we can associate to 
possible stable regions, typical of secular evolution. 
Inside stable -dark- regions, MMRs appear as vertical -light- structures because they generate oscillations in the semimajor axes.
But, inside unstable regions the MMRs appear as dark vertical lines because the planetary perturbations cannot break the small amplitude oscillations typical of resonant motion, specially for strong MMRs. These strong MMRs are clearly defined in all panels and specially in the high eccentricity regime indicating that they provide some stability inside such a chaotic region.
The resonances that persist in several panels of Figure \ref{deltasem338} are: 1:1J at 5.2 au, 2:3J at 6.8 au, 1:2J at 8.2 au, 1:3J at 10.8 au and 1:4J at 13.1 au. At 15.2 au there are two resonances: at low eccentricities the  1:2 with Saturn dominates while at high eccentricities the resonance that dominates is 1:5 with Jupiter. This can be checked calculating the resonance's strength using the code from \citet{2018ascl.soft08002G}.
In the panel corresponding to $e=0.7$ several resonances of the type 1:N with Jupiter are evident up to 1:15 at 31.6 au.

It is interesting to note in the first panel of Figure \ref{deltasem338} the stability of coorbitals of each of the giant planets for a large range of inclinations and the unstable region that surround them.
Note that the coorbitals of Jupiter survive for a wide range of eccentricities while the coorbitals of the other giant planets become unstable for $e>0.2$.
In the subsequent panels, as the eccentricity grows, two diffuse yellow vertical structures -representing very unstable zones- shift from left to right, to larger semimajor axes. They correspond to orbits crossing the orbits of  Jupiter and Saturn. In the last two panels only survives the very wide structure due to Jupiter. 

In the firsts panels wide regions in inclination persist as stable zones but, as the eccentricity grows, unstable regions grow and the only stable regions are limited to niches corresponding to retrograde orbits, see for example panels corresponding to eccentricity 0.5, 0.6 and 0.7.
In particular,
a remarkable feature is evident in all panels, but specially for $e>0.1$: an horizontal stable band defined by $i \sim 150\degr \pm 20\degr$
which seems to be the most stable orbital inclination for a wide range of eccentricities. The observed population  in Figure \ref{obsretro} approximately is concentrated in this stable band.

\begin{figure}
	\centering
	\includegraphics[width=1.0\linewidth]{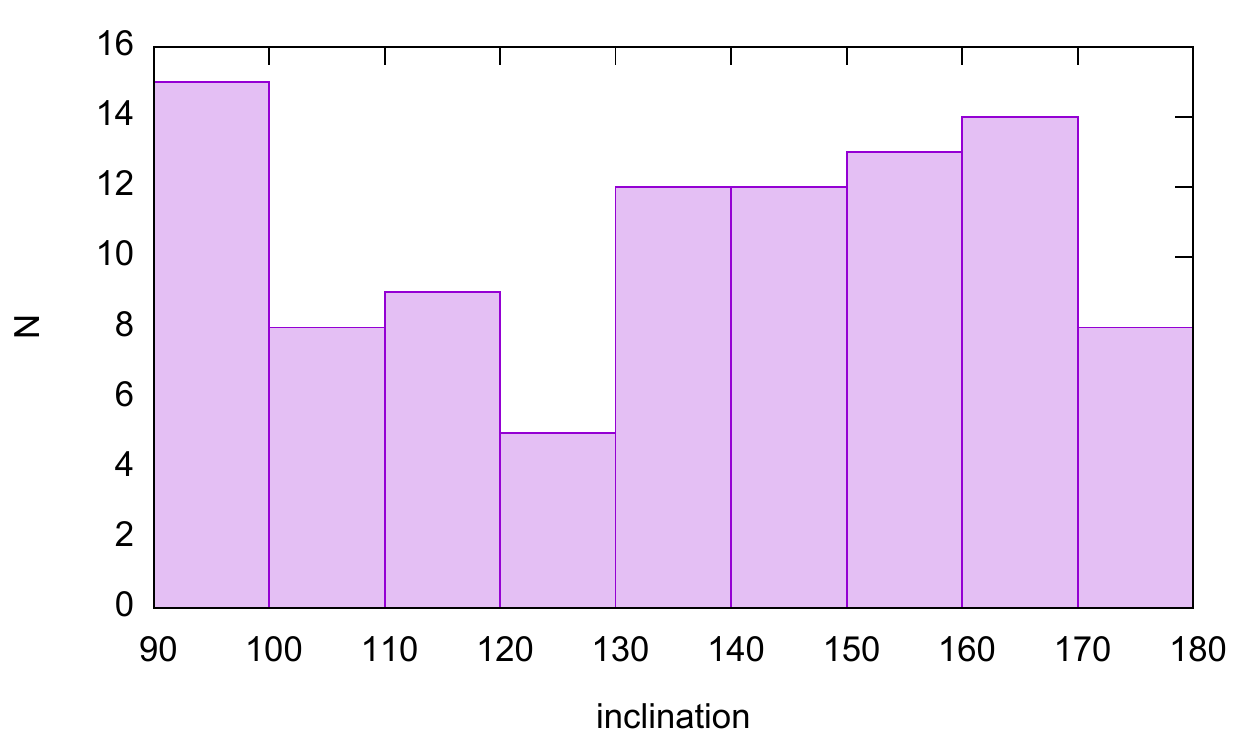}
	\caption{Histogram of the inclination of the retrograde asteroids and comets with $a<40$ au (same sample than Figure \ref{obsretro}). Note the excess from 130 to 170 degrees. }
	\label{histo}
\end{figure}

\begin{figure}
	\centering
	\includegraphics[width=1.0\linewidth]{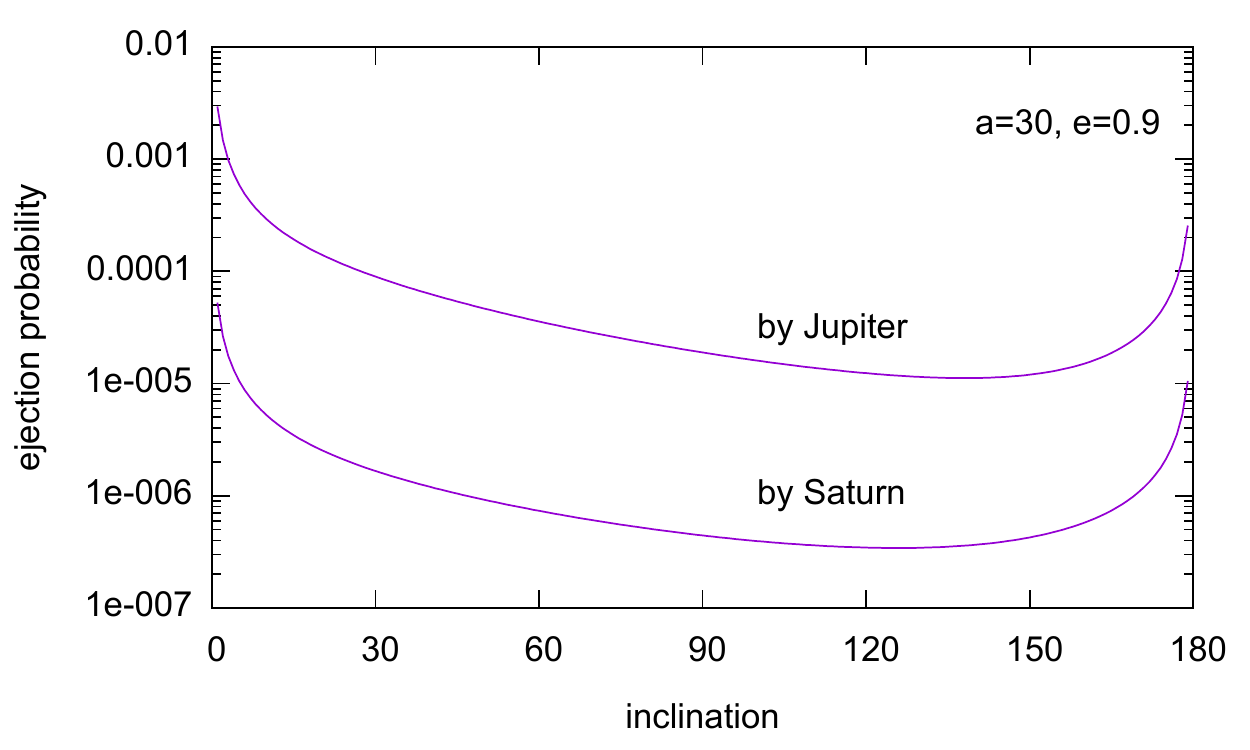}
	\caption{Ejection probability by Jupiter and Saturn for a test particle with $a=30$ au and $e=0.9$ and different inclinations according to \"{O}pik scheme.}
	\label{opik}
\end{figure}

In order to appreciate this concentration we show an histogram of the distribution of the inclinations in Figure
\ref{histo}. We performed a KS test on this distribution and found that it does not belong to a uniform distribution with a probability greater than 0.99.
One of the objects located in this particularly stable stripe is 2015 BZ509, a retrograde coorbital with Jupiter.
This stable band can be understood in terms of the \"{O}pik theory of encounters, 
which is a simplified model based on the geometry of the problem that takes into account the probability of the encounter and also the expected change in the orbital elements due to the encounter.
Just for illustration, following \cite{Valsecchi2000}  we calculated the ejection probability per orbital period for a test particle with $a=30$ au and $e=0.9$ for all inclinations due to the encounters with Jupiter and Saturn. Figure \ref{opik} shows the results where we can see that for the referred interval of inclinations the probability is minimum, then in that interval the planetary perturbations are in general less relevant.

The other evident stable niches that appear specially en the last panels are, as we have stated above, the exterior MMRs of the type 1:N with Jupiter.
These last panels of Figure \ref{deltasem338} can be compared with the figure 7 in \citet{2016MNRAS.461.3075F} which shows that captures in exterior resonances 1:N are very frequent for particles with retrograde orbits for $e\gtrsim 0.75$. That figure shows that in particular for $10 \lesssim a \lesssim 22$ au captures in resonance occur for $40\degr \lesssim i \lesssim 160\degr$ but for $a \gtrsim 22$ au captures only happen for retrograde orbits. Note that 
these limits match very well 
with the most stable regions showed in the lower panels of our Figure \ref{deltasem338}.
Then, the captures found by  these authors  correspond to strong resonances 1:N inside the less
unstable regions of the Solar System for that high eccentricity regime where planetary perturbations are particularly strong.

\begin{figure*}
	\centering
	\includegraphics[width=1.0\linewidth]{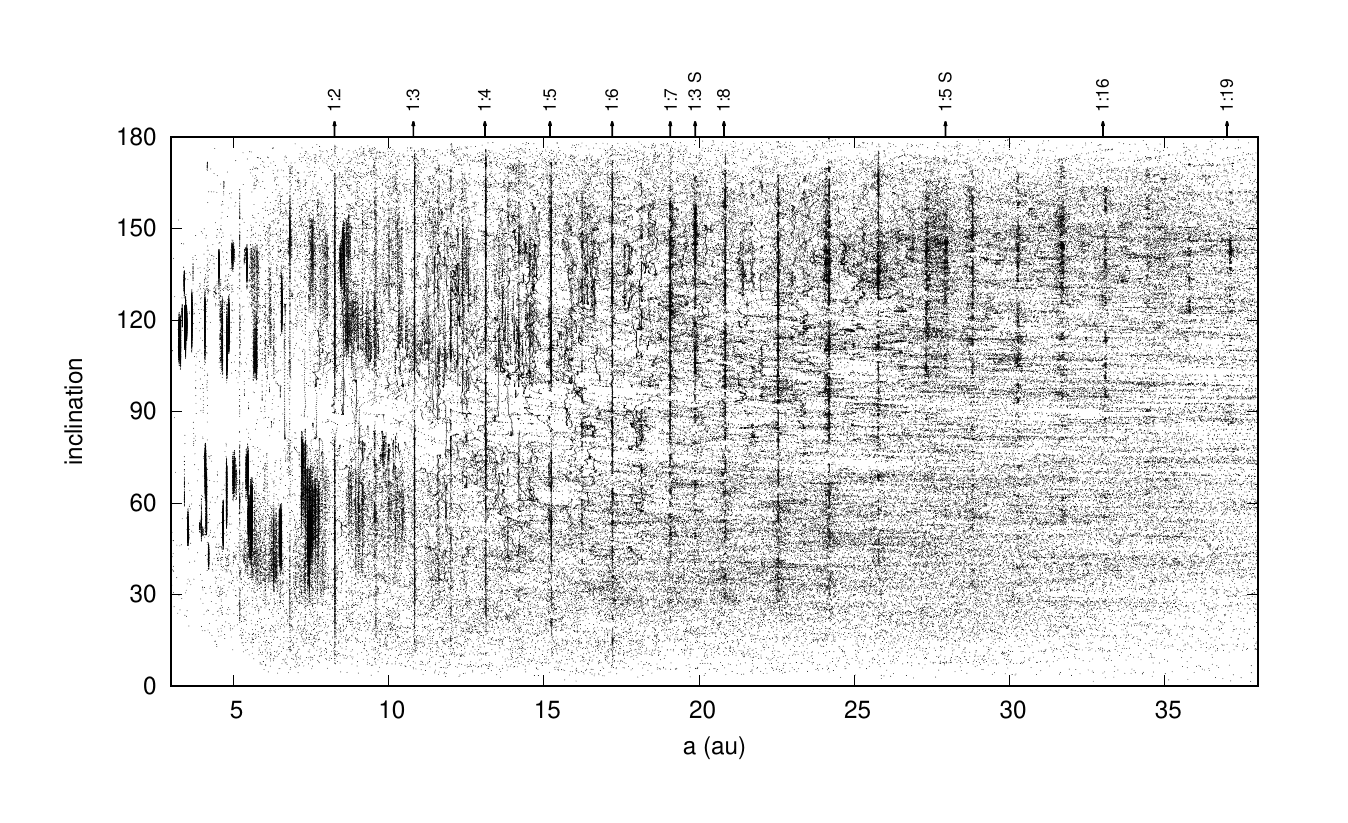}
	\caption{Mean orbital elements verifying  $0.85<e<0.95$ from a numerical integration of 2 Myr of a fictitious population of 1000 test particles. Mean elements calculated using a window of 5000 yrs.  Several resonances of the type 1:N with Jupiter appear as vertical lines and someones are labeled for reference. The resonances 1:3 and 1:5 with Saturn are also detected. Compare with Figure \ref{deltasem338} last panel. }
	\label{mean}
\end{figure*}

\begin{figure}
	\centering
	\includegraphics[width=1.0\linewidth]{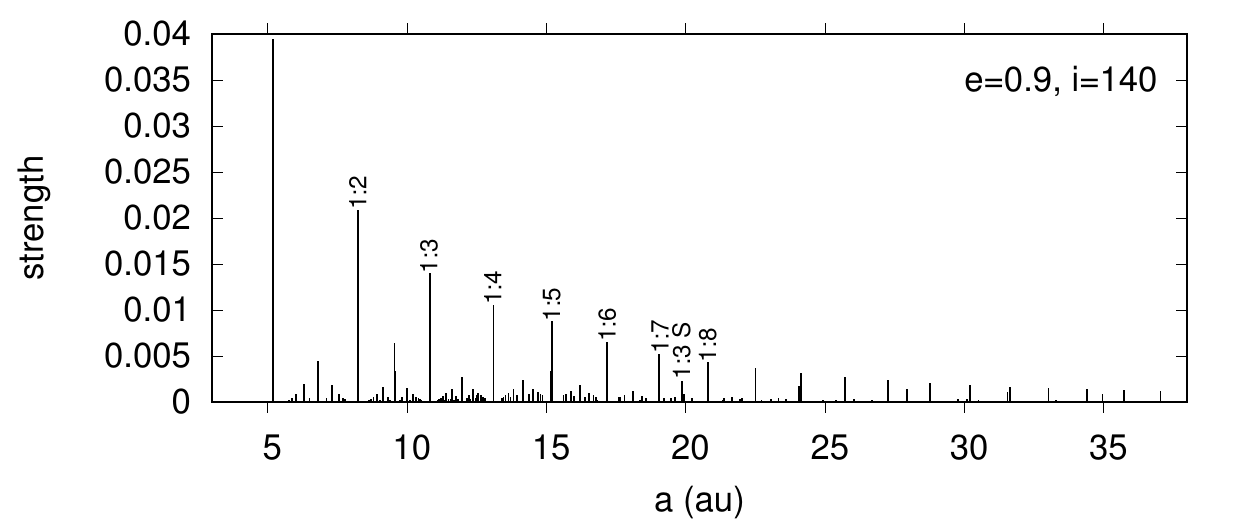}
	\caption{Atlas of the strongest resonances between 5 and 38 au calculated for $e=0.9$, $i=140\degr$ and $\omega = 60\degr$. Resonances of the type 1:N with Jupiter dominate. Some resonances with Jupiter and the resonance 1:3 with Saturn are labeled. Compare with Figure \ref{mean}.}
	\label{atlas}
\end{figure}

The figure by \citet{2016MNRAS.461.3075F} was obtained starting with the sample of observed objects, then the orbital elements were not taken at random from an uniform distribution.
In order to link our dynamical maps with the efficiency of the resonances in capturing particles we have numerically integrated 1000 fictitious particles with orbital elements 
taken at random between the same limits of our maps but all them with initial $e=0.9$. We integrated by 2 Myr using EVORB with output for every 100 yrs and then we have calculated the mean elements using a moving window of 5000 yrs. Figure  \ref{mean} shows the mean orbital elements  plotted for intervals of 1000 yrs satisfying  the condition  $0.85<e<0.95$ in order to compare with Figure \ref{deltasem338} last panel. Note the strong diffusion in semimajor axis for direct orbits which is only halted by resonances located in $a<25$ au. All resonances of the type 1:N with Jupiter can be identified up to 1:19 and also the resonances 1:3 and 1:5 with Saturn. It is evident that in the yellow regions  of Figure  \ref{deltasem338} last panel the resonant motion can not subsist due to the planetary perturbations. Inside that  region the diffusion in semimajor axis is so fast that the resonances cannot imprint any trace. Figure \ref{atlas} shows all resonances with their corresponding strengths calculated according to  \citet{2006Icar..184...29G,2019Icar..317..121G} (see also \url{http://www.fisica.edu.uy/~gallardo/atlas/2bmmr.html}) where the strengths are related to the semiamplitude of the resonant disturbing functions. Resonances of the type 1:N with Jupiter are the strongest even 
for heliocentric distances in between the other giant planets.

Figure \ref{deltasem338}  allows us also to retrieve some of the numerical results by \citet{2015MNRAS.446.1998N,2017MNRAS.467.2673N}. Their study, focused in resonance capture at arbitrary inclinations, was done in quite different scenario with respect to our study because they considered only one planet.  
They concluded that captures in retrograde resonances are more probable  than in direct ones and they argue that the main reason is the 
larger perturbations due to  the planet that a direct object must overcome with respect to a retrograde one. They refer to the perturbations generated by the same planet that generates the resonances but, in the light of our results, we can generalize and assert that the whole outer planetary system is responsible for the destabilization of the direct orbits more than for the retrograde ones.

Then, in the outer solar system several resonances can show up overcoming the planetary perturbations but for extreme eccentricities only the most stronger resonances survive and they are mainly the exterior resonances 1:N with Jupiter.
For the same reason it is not surprising that 1:N resonances with Neptune dominate the trans-Neptunian region \citep{2006Icar..184...29G,2007Icar..192..238L,2018AJ....155..260V}.

\begin{figure}
	\centering
	\includegraphics[width=1.0\linewidth]{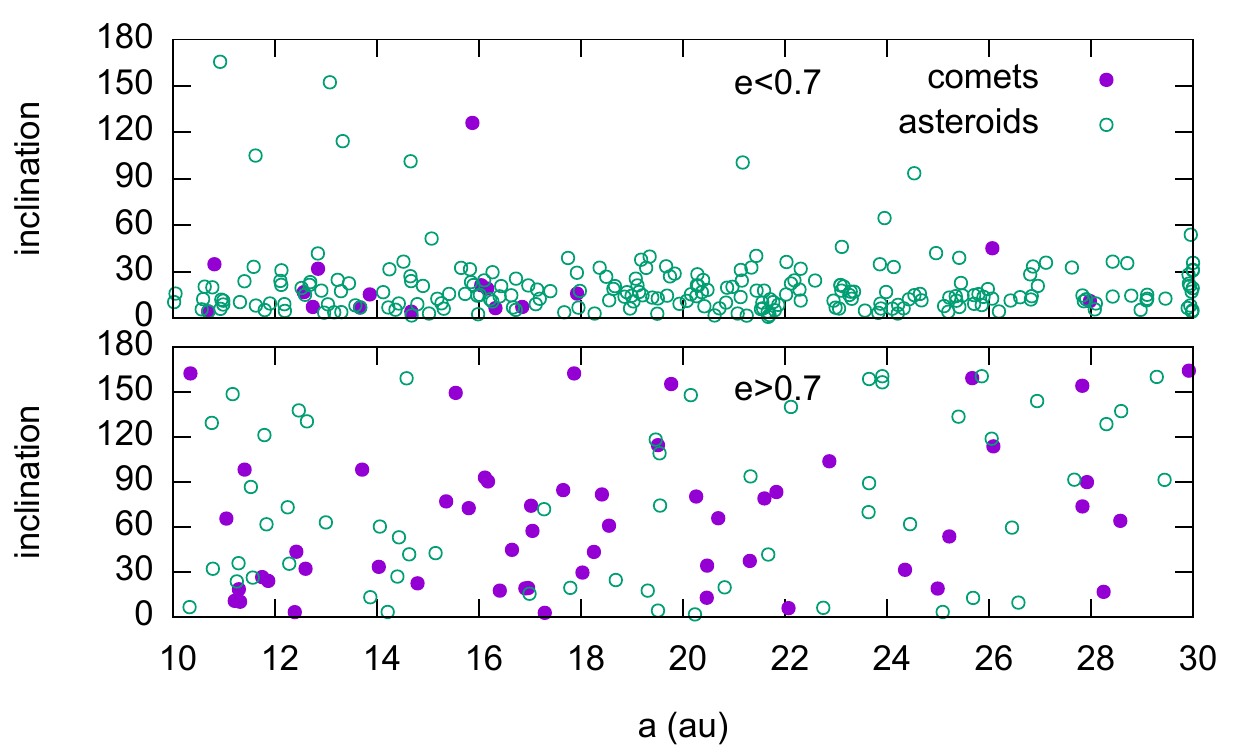}
	\caption{The known population of 367 objects with $10<a<30$ au. Note the change in the distribution of the inclination for $e\sim 0.7$. Source: MPC, February 2019.}
	\label{pobei}
\end{figure}

Figure \ref{pobei}  shows
the known population of 367 objects (297 asteroids and 70 comets) with $10<a<30$ au obtained from MPC\footnote{\url{https://www.minorplanetcenter.net}} database as presented in 
February 2019.
It is remarkable the difference in the inclination's distribution according to the eccentricity. The population with $e<0.7$ is concentrated at low inclinations (mainly $i<30\degr$) while 
the population with  $e>0.7$ is dispersed in all range of inclinations. If we look at Figure \ref{deltasem338} we can verify that this population with $10<a<30$ au is occupying the most unstable regions: for $e<0.7$ is concentrated at the low inclination unstable regions (up to panel corresponding to $e=0.6$) while for $e>0.7$ the population is dispersed in all inclinations inside the yellow regions occupied by the unstable orbits of the last 3 panels of Figure  \ref{deltasem338}.
Then, our maps suggest that the population of objects with $10<a<30$ is mostly the result of the dynamics imposed by the strongest planetary perturbations. 
This is not an extraordinary result because we are dealing mainly with Centaurs for which it is well known they are mostly evolving encountering the giant planets \citep{2004MNRAS.354..798H,2007Icar..190..224D,2009Icar..203..155B,FERNANDEZ20186}.

\citet{FERNANDEZ20186}, studying  the dynamical differences between active and inactive Centaurs, they found that inactive Centaurs have greater dynamical lifetimes than active ones. They argue that longer dynamical lifetimes imply more time exposed to radiation and in consequence the population becomes inactive by depletion of volatiles. 
Probably we are seeing the same effect in the superior panel  of our Figure \ref{pobei}: the bodies located at high inclinations (then, more stable orbits for $e<0.7$) are almost all classified as asteroids, that means, inactive objects.

\section{Dynamical maps for the inner Solar System}
\label{mapsiss}

\begin{figure*}
	\centering
	\includegraphics[width=0.9\linewidth]{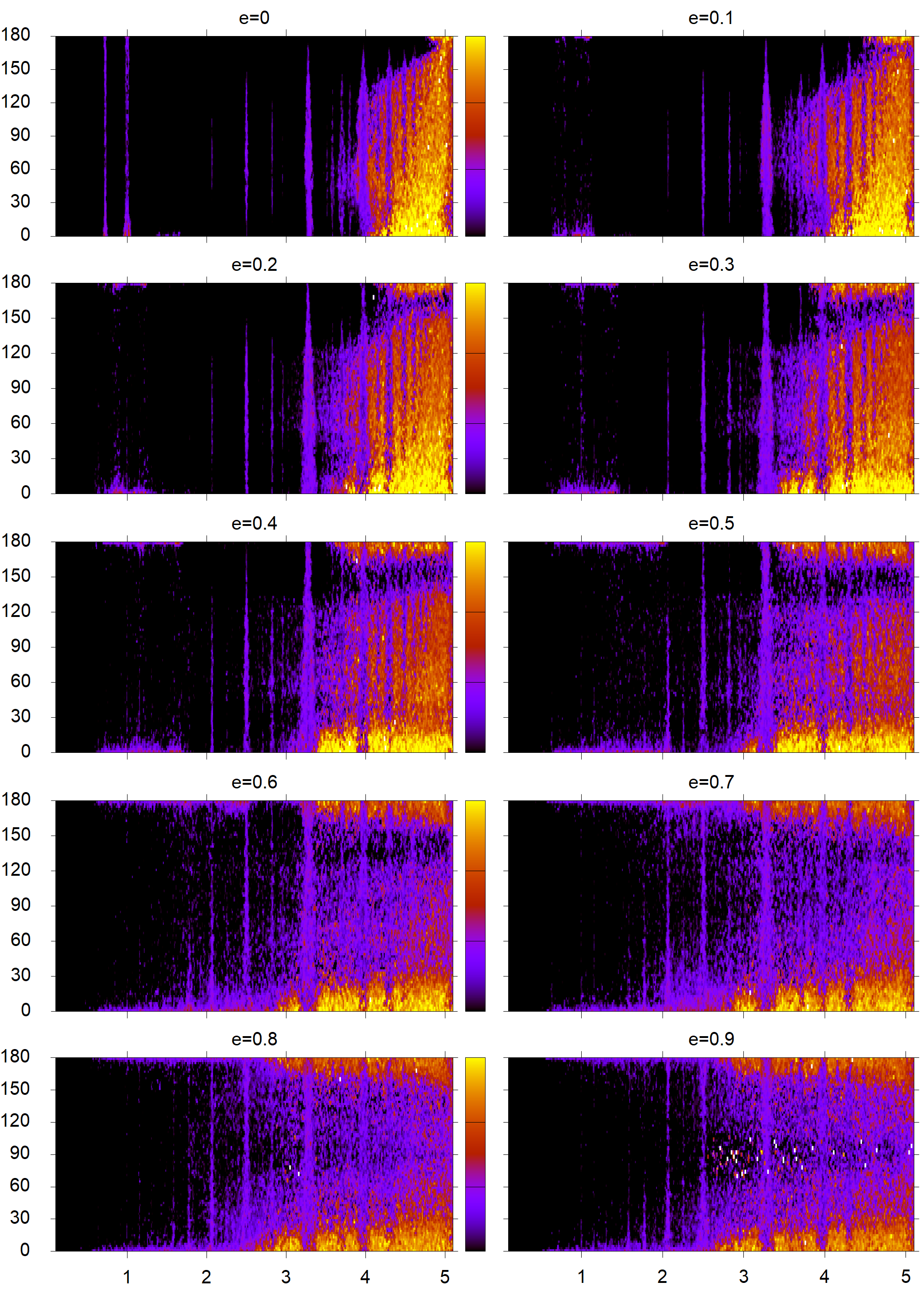}
	\caption{Dynamical maps in terms of initial $(a,i)$ showing the diffusion in $a$ for test particles in the Solar System with initial random $\omega,\Omega,M$ and different initial eccentricities.  Horizontal axis is initial $a$ in au and vertical axis is initial inclination in degrees. 				Going from top (yellow)  to bottom (black), the ticmarks of the logarithmic colour scale indicate 
 	$(\Delta a) /a = 1, 0.1, 0.01$ and $0.001$. White pixels correspond to ejections. The dark vertical bands surrounded by chaotic regions that can be recognized in several panels are generated by stable MMRs. Note that for $a>3.8$ au the low eccentricity orbits are more unstable than high eccentricity ones. For the construction of each map, 45000 test particles per 1000 orbital periods were integrated.
 }
	\label{deltasem05}
\end{figure*}

We constructed dynamical maps for the region $0.1<a<5.1$ au following the same procedure  than in section \ref{mapsoss} but in this case we included also the planets Venus, Earth and Mars. We did not include Mercury due to its small mass and in order to accelerate the numerical integrations. Results  are showed in Figure \ref{deltasem05} using the same scale than in Figure \ref{deltasem338}. In this case
for the construction of each of the maps we integrated 45000 test particles.

MMRs inside stable regions appear as lighter vertical lines and inside unstable regions appear as vertical dark lines as in Figure \ref{deltasem338}.
The main MMRs identified are: the weak coorbitals 1:1V at 0.72 au and
1:1E at 1.0 au, the  4:1J at 2.06 au and the well known 3:1J at  2.50 au, the prominent 2:1J   at 3.28 au present in all panels, the Hilda's resonance 3:2J at 3.97 au and 4:3J at 4.29 au.

\begin{figure}
	\centering
	\includegraphics[width=1.0\linewidth]{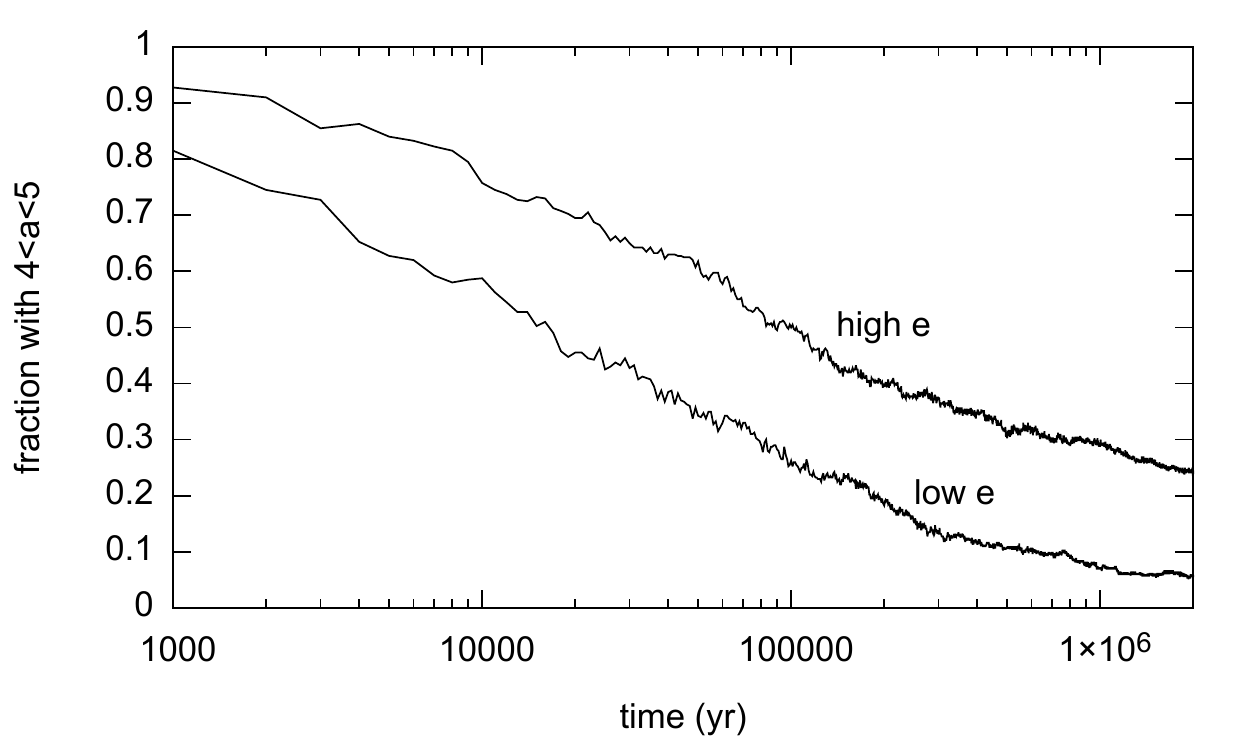}
	\caption{Time evolution of the fraction of survivors from two populations of test particles with initial $4<a<5$, $0\degr<i<180\degr$ and with different  initial eccentricity: one population with $0.05<e<0.15$ and the other with $0.75<e<0.85$.}
	\label{entre4y5}
\end{figure}

The stable band around $i\sim 150\degr$ is also present up to $e\sim 0.7$ but the most interesting result of  Figure \ref{deltasem05} is that
contrary to Figure \ref{deltasem338} and against what we would have expected the 
unstable regions are larger for\textit{ lower} eccentricities.
Compare for example the panel for $e=0.1$ with panel for  $e=0.7$.
This unexpected behavior operates mainly  in the
region $3.8\lesssim a \lesssim 5$ au. We tested this result with numerical integrations of two synthetic populations, each one with 400 test particles
with initial random values taken uniformly distributed in $4<a<5$ au, $0\degr<i<180\degr$ with $\omega$, $\Omega$ and mean anomaly $M$ in the interval $(0\degr,360\degr)$ and different initial eccentricity: one population with $0.05<e<0.15$ and the other with $0.75<e<0.85$.
In
Figure \ref{entre4y5}
we show the time evolution of the fraction of survivors that remained with
 $4<a<5$ au
 from the two populations.
The high eccentricity population between $4<a<5$ au is \textit{more stable} than the low eccentricity population. The half life of the populations differ by a factor of 7. The reason can be found looking at the history of close encounters that the particles had. While the high eccentricity sample had a median of 70 encounters inside Hill's radius with the giant planets  (and 3 with terrestrial planets) the low eccentricity sample had a median of 150 close encounters (and 0 with terrestrial planets). Then, this last sample is substantially more perturbed by the giant planets, mainly Jupiter which is the responsible for the 83 \% of the encounters for the high eccentricity population and the  96 \% of the encounters for the low eccentricity population.

\begin{figure}
	\centering
	\includegraphics[width=1.0\linewidth]{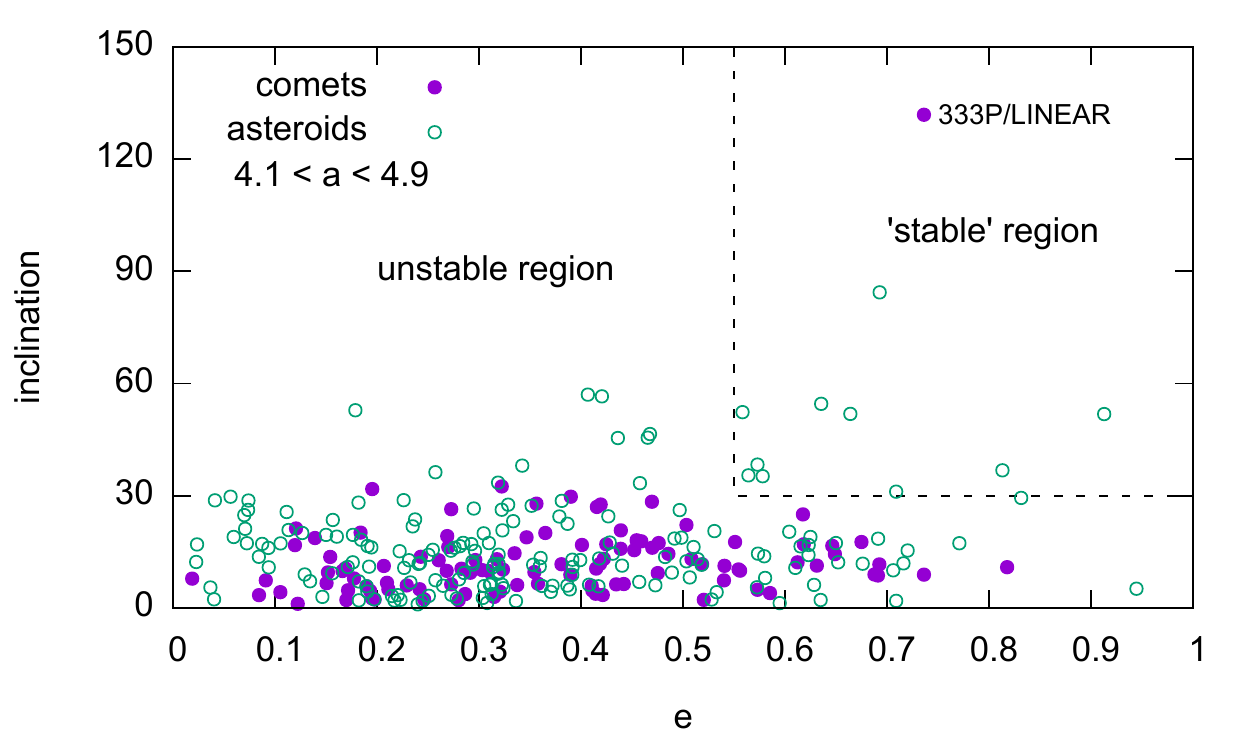}
	\caption{The observed population of asteroids and comets with $4.1<a<4.9$ au.  The most stable region according to Figure \ref{deltasem05} is indicated.   Source: MPC, February 2019.}
	\label{pob4y5}
\end{figure}

This reveals a basic difference in the dynamics of the main belt of asteroids and the region close to Jupiter; while in the main belt an increase in eccentricity favors the encounters with the planets destabilizing the orbits, in the region outside the Hildas an increase in the eccentricity diminishes the frequency of encounters with Jupiter.
These differences in stability have imprints in the distribution of the known population of objects between the asteroid belt and Jupiter.
In
Figure \ref{pob4y5} we plotted all known objects with  $4.1<a<4.9$ au, where the limits were selected in order to avoid 
contamination by Hildas and Trojans. 
  The most stable region between the limits  $4.1<a<4.9$ au   is indicated by a rectangle in the figure
 and it is almost depleted of comets, the only exception is 333P/LINEAR.
Being the high eccentricity region more stable than de low eccentricity one, the objects occupying that region will dynamically survive to more numerous and closer perihelion passages than the low eccentricity population. Then, 
if there were comets at present evolving inside that region, they would be mostly deactivated Jupiter Family comets. In other words, some of the objects shown in the stable region of Figure \ref{pob4y5} could be deactivated comets.
In fact, the asteroids inside the stable region are:
2000 AC229,
2010 OA101,
2012 UU27,
2012 US136,
2013 TA80,
2013 VX9,
2016 EB157,
2016 RR20,
2017 OY68 and
3552 Don Quixote, all of them considered as asteroids in cometary orbits in the literature\footnote{\url{https://www.physics.ucf.edu/~yfernandez/lowtj.html}} \citep{2005AJ....130..308F,Tancredi2014b}.

\section{Conclusions}
\label{conc}

We have characterized the short term orbital stability in the Solar System for the full range of  eccentricities  and inclinations in the region $a<38$ au. Regions in the  space $(a,e,i)$ 
where the destabilizing perturbations by the planets are more important are identified as well as regions where these effects are minimal. In particular we characterized well defined regions where MMRs with the giant planets overcome the other planetary perturbations allowing the existence of protected resonant orbits. Exterior resonances of the type 1:N with Jupiter are clearly present, in particular for retrograde orbits at high eccentricity regime.
Capture in MMR is favored by resonance's strength and impeded by the planetary perturbations in such a way that the competition between both defines the resulting dynamics.

All along the Solar System there is a stripe of approximately 30 degrees in inclination centered around $i \sim 150\degr$ where the planetary perturbations have lower dynamical effects. A concentration around that inclination is observed in the known sample of retrograde small bodies.

Analyzing the distribution of eccentricities and inclinations of the population of small bodies with $10<a<30$ au (mainly Centaurs) we conclude that it is mostly distributed in the unstable regions of ours maps in agreement with the established idea that in general these objects are evolving under strong planetary perturbations.

Our maps for the inner Solar System show 
that in the region between approximately 3.8 and 5 au  the high eccentricity orbits are largely more stable than the low eccentricity ones, contrary to what one might expect. 
This behavior is due to the frequency of close encounters with Jupiter.
Some asteroids evolving in that high eccentricity stable region could be deactivated comets that have lost their volatiles after a long lifetime evolving in high eccentricity (low perihelion) orbits.

\section*{Acknowledgements}

The author thanks the reviewer William Newman who has contributed to improve the original manuscript.
Support from the Comisi\'on Sectorial de Investigaci\'on Cient\'ifica
(CSIC) of the University of the Republic through the project CSIC Grupo I+D
831725 - Planetary Sciences and PEDECIBA are also acknowledged.




\bibliographystyle{mnras}
\bibliography{bibliomapping} 

\begin{thebibliography}{}
\makeatletter
\relax
\def\mn@urlcharsother{\let\do\@makeother \do\$\do\&\do\#\do\^\do\_\do\%\do\~}
\def\mn@doi{\begingroup\mn@urlcharsother \@ifnextchar [ {\mn@doi@}
  {\mn@doi@[]}}
\def\mn@doi@[#1]#2{\def\@tempa{#1}\ifx\@tempa\@empty \href
  {http://dx.doi.org/#2} {doi:#2}\else \href {http://dx.doi.org/#2} {#1}\fi
  \endgroup}
\def\mn@eprint#1#2{\mn@eprint@#1:#2::\@nil}
\def\mn@eprint@arXiv#1{\href {http://arxiv.org/abs/#1} {{\tt arXiv:#1}}}
\def\mn@eprint@dblp#1{\href {http://dblp.uni-trier.de/rec/bibtex/#1.xml}
  {dblp:#1}}
\def\mn@eprint@#1:#2:#3:#4\@nil{\def\@tempa {#1}\def\@tempb {#2}\def\@tempc
  {#3}\ifx \@tempc \@empty \let \@tempc \@tempb \let \@tempb \@tempa \fi \ifx
  \@tempb \@empty \def\@tempb {arXiv}\fi \@ifundefined
  {mn@eprint@\@tempb}{\@tempb:\@tempc}{\expandafter \expandafter \csname
  mn@eprint@\@tempb\endcsname \expandafter{\@tempc}}}

\bibitem[\protect\citeauthoryear{{Bailey} \& {Malhotra}}{{Bailey} \&
  {Malhotra}}{2009}]{2009Icar..203..155B}
{Bailey} B.~L.,  {Malhotra} R.,  2009, \mn@doi [\icarus]
  {10.1016/j.icarus.2009.03.044}, \href
  {http://adsabs.harvard.edu/abs/2009Icar..203..155B} {203, 155}

\bibitem[\protect\citeauthoryear{{Di Sisto} \& {Brunini}}{{Di Sisto} \&
  {Brunini}}{2007}]{2007Icar..190..224D}
{Di Sisto} R.~P.,  {Brunini} A.,  2007, \mn@doi [\icarus]
  {10.1016/j.icarus.2007.02.012}, \href
  {http://adsabs.harvard.edu/abs/2007Icar..190..224D} {190, 224}

\bibitem[\protect\citeauthoryear{{Duncan}, {Levison}  \& {Budd}}{{Duncan}
  et~al.}{1995}]{1995AJ....110.3073D}
{Duncan} M.~J.,  {Levison} H.~F.,   {Budd} S.~M.,  1995, \mn@doi [\aj]
  {10.1086/117748}, \href {http://adsabs.harvard.edu/abs/1995AJ....110.3073D}
  {110, 3073}

\bibitem[\protect\citeauthoryear{{Evans} \& {Tabachnik}}{{Evans} \&
  {Tabachnik}}{1999}]{1999Natur.399...41E}
{Evans} N.~W.,  {Tabachnik} S.,  1999, \mn@doi [\nat] {10.1038/19919}, \href
  {http://adsabs.harvard.edu/abs/1999Natur.399...41E} {399, 41}

\bibitem[\protect\citeauthoryear{{Fern{\'a}ndez}}{{Fern{\'a}ndez}}{2005}]{2005ASSL..328.....F}
{Fern{\'a}ndez} J.~A.,  ed. 2005, {Comets - Nature, Dynamics, Origin and their
  Cosmological Relevance}  Astrophysics and Space Science Library Vol. 328,
  \mn@doi{10.1007/978-1-4020-3495-4.
}

\bibitem[\protect\citeauthoryear{{Fern{\'a}ndez}, {Gallardo}  \&
  {Brunini}}{{Fern{\'a}ndez} et~al.}{2002}]{2002Icar..159..358F}
{Fern{\'a}ndez} J.~A.,  {Gallardo} T.,   {Brunini} A.,  2002, \mn@doi [\icarus]
  {10.1006/icar.2002.6903}, \href
  {http://adsabs.harvard.edu/abs/2002Icar..159..358F} {159, 358}

\bibitem[\protect\citeauthoryear{{Fern{\'a}ndez}, {Jewitt}  \&
  {Sheppard}}{{Fern{\'a}ndez} et~al.}{2005}]{2005AJ....130..308F}
{Fern{\'a}ndez} Y.~R.,  {Jewitt} D.~C.,   {Sheppard} S.~S.,  2005, \mn@doi
  [\aj] {10.1086/430802}, \href
  {http://adsabs.harvard.edu/abs/2005AJ....130..308F} {130, 308}

\bibitem[\protect\citeauthoryear{{Fern{\'a}ndez}, {Gallardo}  \&
  {Young}}{{Fern{\'a}ndez} et~al.}{2016}]{2016MNRAS.461.3075F}
{Fern{\'a}ndez} J.~A.,  {Gallardo} T.,   {Young} J.~D.,  2016, \mn@doi [\mnras]
  {10.1093/mnras/stw1532}, \href
  {http://adsabs.harvard.edu/abs/2016MNRAS.461.3075F} {461, 3075}

\bibitem[\protect\citeauthoryear{Fernandez, Helal  \& Gallardo}{Fernandez
  et~al.}{2018}]{FERNANDEZ20186}
Fernandez J.~A.,  Helal M.,   Gallardo T.,  2018, \mn@doi [Planetary and Space
  Science] {https://doi.org/10.1016/j.pss.2018.05.013}, 158, 6

\bibitem[\protect\citeauthoryear{{Gallardo}}{{Gallardo}}{2006}]{2006Icar..184...29G}
{Gallardo} T.,  2006, \mn@doi [\icarus] {10.1016/j.icarus.2006.04.001}, \href
  {http://adsabs.harvard.edu/abs/2006Icar..184...29G} {184, 29}

\bibitem[\protect\citeauthoryear{{Gallardo}}{{Gallardo}}{2018}]{2018ascl.soft08002G}
{Gallardo} T.,  2018, {rsigma: Resonant disturbance}, Astrophysics Source Code
  Library (\mn@eprint {ascl} {1808.002})

\bibitem[\protect\citeauthoryear{{Gallardo}}{{Gallardo}}{2019}]{2019Icar..317..121G}
{Gallardo} T.,  2019, \mn@doi [\icarus] {10.1016/j.icarus.2018.07.002}, \href
  {https://ui.adsabs.harvard.edu/\#abs/2019Icar..317..121G} {317, 121}

\bibitem[\protect\citeauthoryear{{Gladman} \& {Duncan}}{{Gladman} \&
  {Duncan}}{1990}]{1990AJ....100.1680G}
{Gladman} B.,  {Duncan} M.,  1990, \mn@doi [\aj] {10.1086/115628}, \href
  {http://adsabs.harvard.edu/abs/1990AJ....100.1680G} {100, 1680}

\bibitem[\protect\citeauthoryear{{Grazier}, {Newman}, {Kaula}  \&
  {Hyman}}{{Grazier} et~al.}{1999a}]{1999Icar..140..341G}
{Grazier} K.~R.,  {Newman} W.~I.,  {Kaula} W.~M.,   {Hyman} J.~M.,  1999a,
  \mn@doi [\icarus] {10.1006/icar.1999.6146}, \href
  {http://adsabs.harvard.edu/abs/1999Icar..140..341G} {140, 341}

\bibitem[\protect\citeauthoryear{{Grazier}, {Newman}, {Varadi}, {Kaula}  \&
  {Hyman}}{{Grazier} et~al.}{1999b}]{1999Icar..140..353G}
{Grazier} K.~R.,  {Newman} W.~I.,  {Varadi} F.,  {Kaula} W.~M.,   {Hyman}
  J.~M.,  1999b, \mn@doi [\icarus] {10.1006/icar.1999.6147}, \href
  {http://adsabs.harvard.edu/abs/1999Icar..140..353G} {140, 353}

\bibitem[\protect\citeauthoryear{{Holman} \& {Wisdom}}{{Holman} \&
  {Wisdom}}{1993}]{1993AJ....105.1987H}
{Holman} M.~J.,  {Wisdom} J.,  1993, \mn@doi [\aj] {10.1086/116574}, \href
  {http://adsabs.harvard.edu/abs/1993AJ....105.1987H} {105, 1987}

\bibitem[\protect\citeauthoryear{{Horner}, {Evans}  \& {Bailey}}{{Horner}
  et~al.}{2004}]{2004MNRAS.354..798H}
{Horner} J.,  {Evans} N.~W.,   {Bailey} M.~E.,  2004, \mn@doi [\mnras]
  {10.1111/j.1365-2966.2004.08240.x}, \href
  {http://adsabs.harvard.edu/abs/2004MNRAS.354..798H} {354, 798}

\bibitem[\protect\citeauthoryear{{Lykawka} \& {Mukai}}{{Lykawka} \&
  {Mukai}}{2007}]{2007Icar..192..238L}
{Lykawka} P.~S.,  {Mukai} T.,  2007, \mn@doi [\icarus]
  {10.1016/j.icarus.2007.06.007}, \href
  {http://adsabs.harvard.edu/abs/2007Icar..192..238L} {192, 238}

\bibitem[\protect\citeauthoryear{{Morais} \& {Namouni}}{{Morais} \&
  {Namouni}}{2013}]{2013MNRAS.436L..30M}
{Morais} M.~H.~M.,  {Namouni} F.,  2013, \mn@doi [\mnras]
  {10.1093/mnrasl/slt106}, \href
  {http://adsabs.harvard.edu/abs/2013MNRAS.436L..30M} {436, L30}

\bibitem[\protect\citeauthoryear{{Morais} \& {Namouni}}{{Morais} \&
  {Namouni}}{2017}]{2017MNRAS.472L...1M}
{Morais} M.~H.~M.,  {Namouni} F.,  2017, \mn@doi [\mnras]
  {10.1093/mnrasl/slx125}, \href
  {http://adsabs.harvard.edu/abs/2017MNRAS.472L...1M} {472, L1}

\bibitem[\protect\citeauthoryear{{Namouni} \& {Morais}}{{Namouni} \&
  {Morais}}{2015}]{2015MNRAS.446.1998N}
{Namouni} F.,  {Morais} M.~H.~M.,  2015, \mn@doi [\mnras]
  {10.1093/mnras/stu2199}, \href
  {http://adsabs.harvard.edu/abs/2015MNRAS.446.1998N} {446, 1998}

\bibitem[\protect\citeauthoryear{{Namouni} \& {Morais}}{{Namouni} \&
  {Morais}}{2017}]{2017MNRAS.467.2673N}
{Namouni} F.,  {Morais} M.~H.~M.,  2017, \mn@doi [\mnras]
  {10.1093/mnras/stx290}, \href
  {http://adsabs.harvard.edu/abs/2017MNRAS.467.2673N} {467, 2673}

\bibitem[\protect\citeauthoryear{{Robutel} \& {Laskar}}{{Robutel} \&
  {Laskar}}{2001}]{2001Icar..152....4R}
{Robutel} P.,  {Laskar} J.,  2001, \mn@doi [\icarus] {10.1006/icar.2000.6576},
  \href {http://adsabs.harvard.edu/abs/2001Icar..152....4R} {152, 4}

\bibitem[\protect\citeauthoryear{Tancredi}{Tancredi}{2014}]{Tancredi2014b}
Tancredi G.,  2014, \mn@doi [Icarus] {10.1016/j.icarus.2014.02.013}, 234, 66

\bibitem[\protect\citeauthoryear{{Torbett}}{{Torbett}}{1989}]{1989AJ.....98.1477T}
{Torbett} M.~V.,  1989, \mn@doi [\aj] {10.1086/115233}, \href
  {http://adsabs.harvard.edu/abs/1989AJ.....98.1477T} {98, 1477}

\bibitem[\protect\citeauthoryear{{Torbett} \& {Smoluchowski}}{{Torbett} \&
  {Smoluchowski}}{1990}]{1990Natur.345...49T}
{Torbett} M.~V.,  {Smoluchowski} R.,  1990, \mn@doi [\nat] {10.1038/345049a0},
  \href {http://adsabs.harvard.edu/abs/1990Natur.345...49T} {345, 49}

\bibitem[\protect\citeauthoryear{Valsecchi, Milani, Gronchi  \&
  Chesley}{Valsecchi et~al.}{2000}]{Valsecchi2000}
Valsecchi G.,  Milani A.,  Gronchi G.,   Chesley S.,  2000, Celestial Mechanics
  and Dynamical Astronomy, 78, 83

\bibitem[\protect\citeauthoryear{{Volk} et~al.,}{{Volk}
  et~al.}{2018}]{2018AJ....155..260V}
{Volk} K.,  et~al., 2018, \mn@doi [\aj] {10.3847/1538-3881/aac268}, \href
  {http://adsabs.harvard.edu/abs/2018AJ....155..260V} {155, 260}

\bibitem[\protect\citeauthoryear{{Wiegert}, {Connors}  \& {Veillet}}{{Wiegert}
  et~al.}{2017}]{2017Natur.543..687W}
{Wiegert} P.,  {Connors} M.,   {Veillet} C.,  2017, \mn@doi [\nat]
  {10.1038/nature22029}, \href
  {http://adsabs.harvard.edu/abs/2017Natur.543..687W} {543, 687}

\makeatother
\end{thebibliography}




%
%


\bsp	
\label{lastpage}
\end{document}